\documentclass[pre,superscriptaddress,twocolumn,amsmath,amssymb,floatfix,showpacs]{revtex4}

\usepackage{graphicx}

\begin{document}

\title{Spatial patterns of close relationships across the lifespan} 
\author{Hang-Hyun Jo}
\email[Corresponding author: ]{johanghyun@postech.ac.kr}
\affiliation{BECS, Aalto University School of Science, P.O. Box 12200, FI-00076, Finland}
\affiliation{BK21plus Physics Division and Department of Physics, Pohang University of Science and Technology, Pohang 790-784, Republic of Korea}
\author{Jari Saram\"aki}
\affiliation{BECS, Aalto University School of Science, P.O. Box 12200, FI-00076, Finland}
\author{Robin I. M. Dunbar}
\affiliation{Department of Experimental Psychology, University of Oxford, South Parks Road, Oxford OX1 3UD, UK}
\affiliation{BECS, Aalto University School of Science, P.O. Box 12200, FI-00076, Finland}
\author{Kimmo Kaski}
\affiliation{BECS, Aalto University School of Science, P.O. Box 12200, FI-00076, Finland}
\affiliation{Department of Experimental Psychology, University of Oxford, South Parks Road, Oxford OX1 3UD, UK}
\affiliation{CABDyN Complexity Centre, Sa\"id Business School, University of Oxford, Park End Street, Oxford OX1 1HP, UK}
\affiliation{Center for Complex Network Research, Department of Physics, Northeastern University, Boston, MA 02115, USA}

\date{\today}

\begin{abstract}
The dynamics of close relationships is important for understanding the migration patterns of individual life-courses. The bottom-up approach to this subject by social scientists has been limited by sample size, while the more recent top-down approach using large-scale datasets suffers from a lack of detail about the human individuals. We incorporate the geographic and demographic information of millions of mobile phone users with their communication patterns to study the dynamics of close relationships and its effect in their life-course migration. We demonstrate how the close age- and sex-biased dyadic relationships are correlated with the geographic proximity of the pair of individuals, e.g., young couples tend to live further from each other than old couples. In addition, we find that emotionally closer pairs are living geographically closer to each other. These findings imply that the life-course framework is crucial for understanding the complex dynamics of close relationships and their effect on the migration patterns of human individuals.
\end{abstract}



\maketitle

\section{Introduction}

Human societies have successfully been described in the framework of social networks based on dyadic relationships~\cite{Borgatti2009}. In 
recent years, a number of social networks have been characterised in terms of small-world 
properties~\cite{Watts1998}, broad distributions of the number of neighbours~\cite{Barabasi1999}, assortative mixing~\cite{Newman2002} or homophily~\cite{McPherson2001}, and community structure~\cite{Fortunato2010}. This is partially due to recent access to large-scale highly-resolved digital datasets on human dynamics and social interaction~\cite{Lazer2009}. Mobile phone datasets in particular have provided a unique opportunity to study the structure and dynamics of human relationships~\cite{Onnela2007,Onnela2007b,Eagle2009,Eagle2010,miritello2013time,miritello2013limited}. Although the lack of detail about 
individuals 
often undermines the importance of large-scale studies based on anonymised datasets, limited geographic and demographic information of mobile phone users has successfully been used, e.g., in studies of age and sex biases in social relationships~\cite{Palchykov2012,Kovanen2013}.

It is important to stress that humans are embedded not only in social networks but also in 
geographic space~\cite{Lambiotte2008,Barthelemy2011}. People move and migrate for a number of reasons. For example, people leave their parental home for education or employment, to get married, to rear a family, or they can move due to divorce and separation~\cite{Kley2011}. In these life-course events, 
close relationships play a crucial role in shaping 
migrational patterns. Human mobility patterns have recently been studied by using the datasets of time-resolved location of mobile phone users~\cite{Gonzalez2008,Krings2009,Simini2012,Palchykov2014}. These datasets are limited to periods of a few years at most, so far allowing only cross-sectional analysis. In contrast, the longitudinal approach adopted by social scientists has been used to investigate long-term migration patterns over the human lifespan~\cite{Speare1970,Courgeau1990}, 
but suffers from the fact that sample size is invariably limited.

Large-scale mobile phone datasets can be used to understand the role of close relationships in the life-course migration, by 
exploiting geographic and demographic information of mobile phone users. Frequency of contact between a pair of individuals has been established as a reliable index of emotional closeness in relationships, and the frequency of contact by telephone and other digital media like email and text message is known to correlate significantly with the frequency of face-to-face contact~\cite{Roberts2009,Roberts2011,Saramaki2014}. Thus one can assume that most of important relationships of individuals are captured by mobile phone communication records, and that the level of emotional closeness in a relationship is reflected in the strength of communication. We also assume that the life-course migration patterns are reflected in the age- and sex-dependent geographic correlations of users. In other words, even though our mobile phone dataset of users of all age groups is necessarily cross-sectional, we can use them to gain insights into longitudinal, life-course migration patterns.

\section{Results}

We analyse a large-scale mobile phone call dataset from a single mobile service provider in a European country~\cite{Onnela2007,Onnela2007b}. This dataset spans the first 7 months of 2007 and contains around 1.9 billion calls between 33 million individual users. Among them, around 5.1 million users subscribed to the provider, meaning that their demographic information is partially available. 
In order to study the closest relationships of egos (individuals), we rank their alters (friends, family members, acquaintances) in terms of the number of calls to/from the ego, irrespective of whether such alters have the relevant demographic information. Initially we focus on the top-ranked and 2nd-ranked alters. We consider only those ego-alter pairs for which information on age, sex, and geographic location is available for both ego and alter. There are 1.1 million pairs of egos and top-ranked alters 
and 0.7 million pairs of egos and 2nd-ranked alters.

For each ego-alter pair, we determine the geographic difference index $h_{ij}$ that has the value of $0$ ($1$) if the alter $j$ lives in the same (different) municipality as the ego $i$ (see Methods for determination of the municipality). Let us first set the baseline by calculating the average geographic difference indices. For all pairs of egos and top-ranked alters, $\langle h\rangle \approx 0.58$. The average indices for female egos and for male egos are $0.59$ and $0.58$, respectively, i.e., no sex differences are observed.
Here we have considered egos whose number of calls is at least $15$, corresponding to one call per two weeks on average. We like to note that our conclusion is robust against to the variation of this criterion. 

\subsection{Geographic locations of highly-ranked alters and sex differences} 

\begin{figure}[!t]
    \includegraphics[width=\columnwidth]{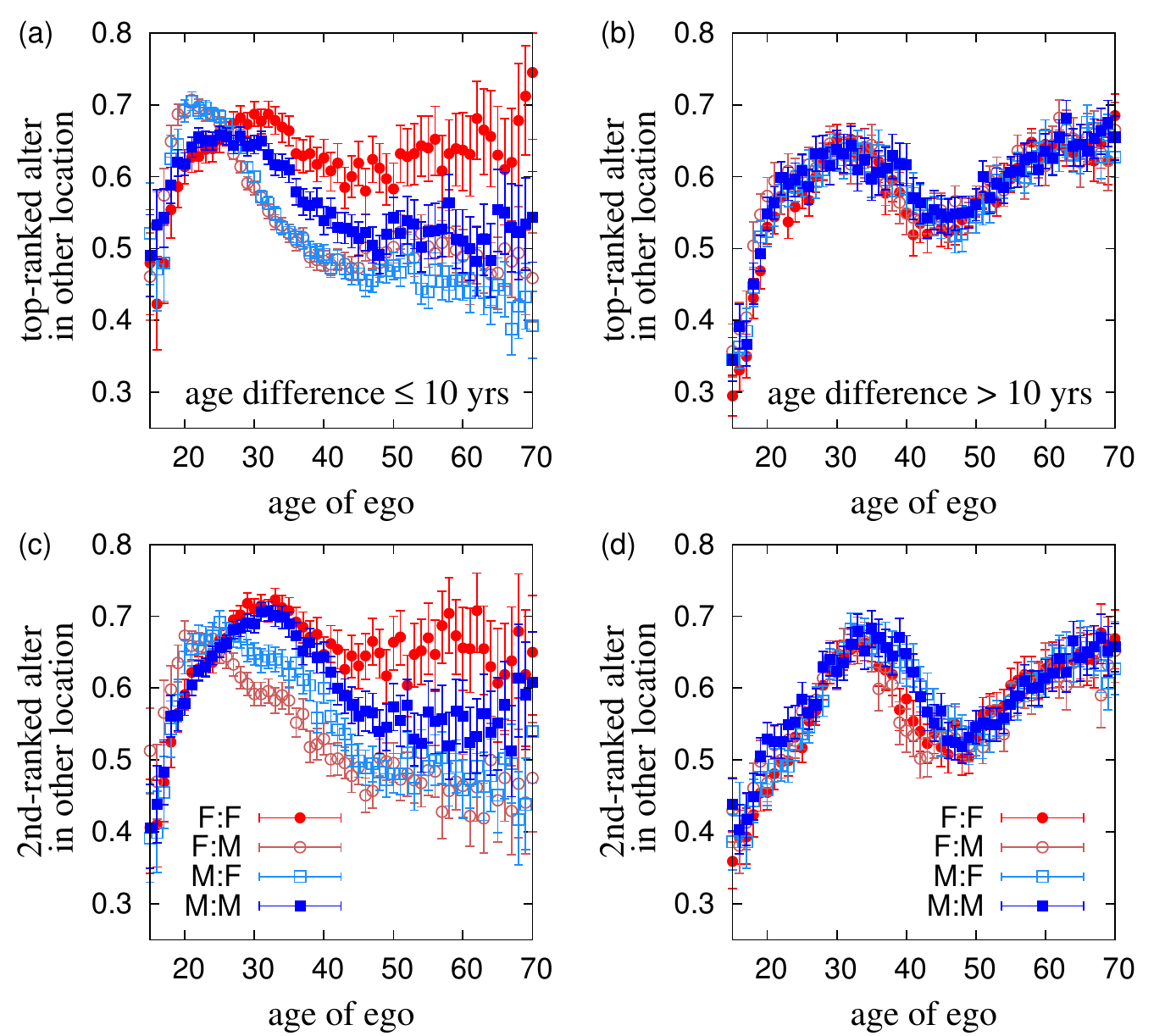}
    \caption{Geographic difference indices of top-ranked and 2nd-ranked alters who live in a different municipality to ego (top and bottom, respectively). The left panels are for ego-alter pairs with age difference less than 10 years, while the age difference is larger than 10 years in the right panels. We use the notational convention that, for example, ``M:F'' denotes male ego and female alter. Error bars show the confidence interval with significance level $\alpha=0.05$.}
\label{fig:allBFZip}
\end{figure}

The variation 
in geographic correlations across 
the lifespan can be revealed by using more detailed information on the egos and alters. We split the ego-alter pairs into four groups based on sex (M for male, F for female): F:F, M:M, F:M, and M:F. We also consider the ages of both egos and alters, and divide ego-alter pairs into two categories where the age difference between egos and alters is $\leq 10$ years and $>10$ years. This division is motivated by the observation that the age distribution of top-ranked alters is bimodal, with one peak approximately around the ego's own age and another $\approx 25$ years apart, roughly corresponding to one generation~\cite{Palchykov2012}. The average geographic difference indices $\langle h \rangle$, i.e., the fractions of alters living in a different municipality to the ego, are shown in Fig.~\ref{fig:allBFZip} as a function of the ego's age for the different sex and age groups. 

For top-ranked alters of the opposite sex (F:M, M:F) with age difference $\leq 10$ years, the geographic difference indices for female and male egos are mostly identical within error bars (Fig.~\ref{fig:allBFZip}(a)). The fraction of alters in a different municipality increases up to $\approx 0.7$ for egos who are around 20 years old, then it gradually decreases to $\approx 0.45$ by the mid-40's, and 
then remains at approximatively the same level thereafter. If one assumes that this behaviour is caused by romantic ego-alter relationships, a possible interpretation is that young couples live in the same neighbourhood until going to work or college necessitates a larger distance; 
eventually, individuals settle down together with either the same or another partner at an older age. The indices for same-sex ego-alter pairs with age difference $\leq 10$ years behave differently and show sex dependence. They increase more slowly, reach the maximum in the late 20's (M:M) or around 30 years (F:F), then decrease and fluctuate (M:M) or slightly decrease and increase again (F:F). After the peak, the M:M curve remains lower than the F:F curve at all ages, indicating that the top-ranked male alters of males live on average geographically closer.

\begin{figure}[!t]
    \includegraphics[width=\columnwidth]{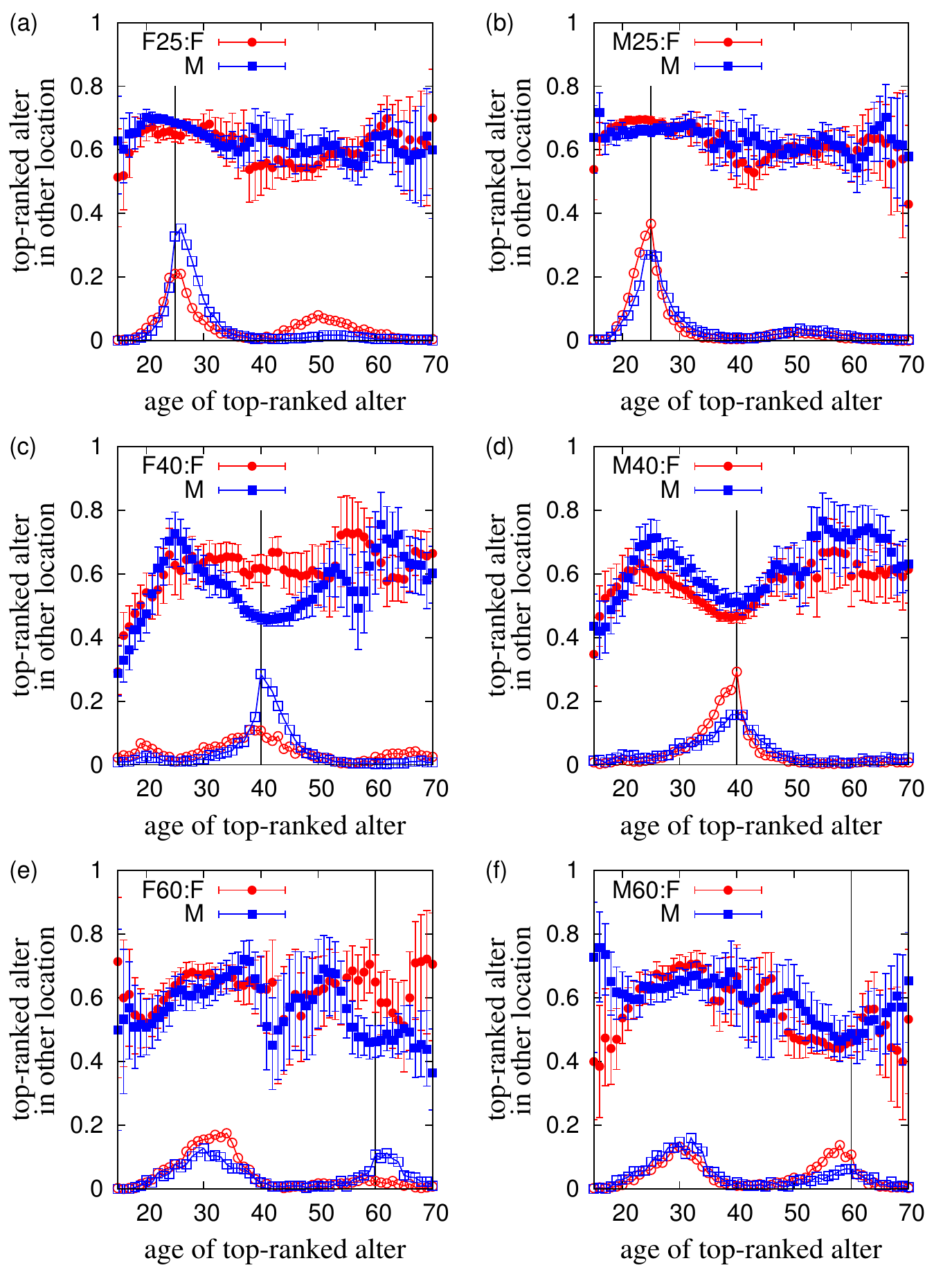}
    \caption{Geographic difference indices of top-ranked alters who live in a different municipality to ego for several demographic groups of egos of ages 25, 40, and 60 (top to bottom), and of females (left) and males (right). The fraction for each age, $a$, has been averaged over $[a-2,a+2]$ for accentuating trends. Error bars show the confidence interval with significance level $\alpha=0.05$. In each panel, lower curves with empty symbols represent age distributions of top-ranked alters, which are magnified $2.5$ times for clearer visualization. Vertical black lines at the ego age are added for guiding the eye.} 
\label{fig:1BF}
\end{figure}



For top-ranked alters with age difference $>10$ years, the sexes of egos and alters appear to play no role (Fig.~\ref{fig:allBFZip}(b)). For all cases, when the egos are young, $\approx 65\%$ of top-ranked alters live in the same municipality. 
The fraction of alters in different municipality peaks in the 30's, shows a local minimum at around mid-40's, and then increases again. It would be reasonable to expect that this is because children live with their parents until leaving home in their 20's; this is reflected in the local minimum at around mid-40's as well. For older egos, children have left home and live elsewhere, increasing the geographic difference index with ego's age.

\begin{figure}[!t]
    \includegraphics[width=\columnwidth,trim=10mm 0mm 20mm 0mm,clip]{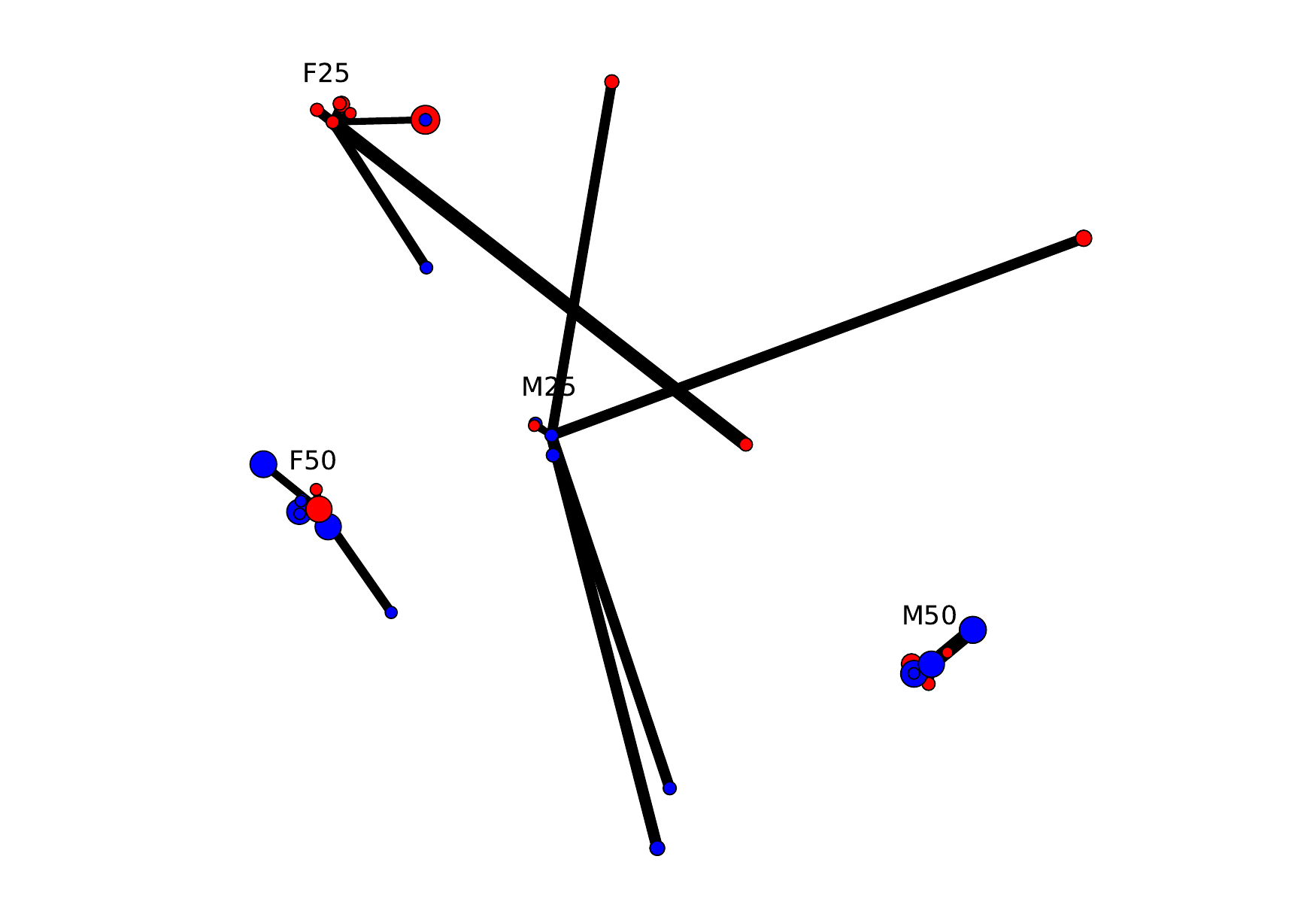}
    \caption{Geospatial layout of egocentric networks for four egos. The red and blue filled circles denote female and male individual users, respectively, and the size of circle is proportional to the age of the user. The width of an edge corresponds to the number of calls between users. Some overlapping circles have been moved a small amount for the sake of clarity.}
\label{fig:egoNets}
\end{figure}


The geographic difference indices between egos and their 2nd-ranked alters behave roughly similarly to the top-ranked alters (Fig.~\ref{fig:allBFZip}(c--d)). The main difference is that the indices peak somewhat later; furthermore, for age difference $\leq 10$ years, the F:M and M:F curves no longer overlap, but the 2nd-ranked alters of females live slightly more often in the same municipality with the egos. There are several possible reasons for this behaviour and no definite conclusions can be drawn from  demographic data alone. 
For example, this behaviour might arise from the partners of female egos being ranked 2nd more often than the partners of male egos. In particular women switch their focal interest from their partners to their children as these become adult; however this would only affect the geographic difference indices for female egos old enough to have grown-up children. 
For younger women, it may relate to the need to have a close female friend in whim to confide; women are more likely to have a same-sex intimate friend than men are. 
For ego-alter pairs with age difference $>10$ years, ego and alter sexes do not matter, except for female egos in their 30's--40's whose 2nd-ranked alters are slightly more often located in the same municipality, compared to corresponding male egos. This effect might be because of mother-child relationships, considering the age difference. 

\subsection{Correlations with alter age} 

These geographic correlations can be more clearly understood by considering age correlations between highly-ranked alters and their geographic difference indices. We show the age distributions of top-ranked alters and their geographic difference indices for several demographic groups of egos in Fig.~\ref{fig:1BF}. We use the notational convention that, for example, ``F25'' denotes the group of 25-year-old female egos. For the F25 and M25 groups (18675 and 21018 pairs, respectively), the curves for geographic difference indices are overall high and relatively flat, showing little correlation between demographic and geographic information. For 40-year-old females (F40, 8550 pairs) (Fig.~\ref{fig:1BF}(c)), the male top-ranked alters are on average slightly older than the egos, and tend to live close to egos; it is reasonable to expect these to be their partners. In contrast, female top-ranked alters of similar age live more often in another municipality. The age distribution of top-ranked alters shows a small peak at around 20 years of age, more pronounced for female egos. As these alters commonly live in the same municipality, this peak can likely be attributed to the children of egos. For the M40 (10933 pairs), the overall pattern is fairly similar. For the F60 and M60 groups (4360 and 5115 pairs, respectively), the data 
display a lot of variation because of smaller 
samples. However, top-ranked alters of the same age tend to be more often found nearby. Additionally, there are clear peaks in the distribution of alter ages at around 30 years old; these can again be attributed to parent-child relationships. This finer scale analysis is comparable to the observed geographic correlations in Fig.~\ref{fig:allBFZip}(a--b).

\subsection{Geography and alter ranks} 

\begin{figure}[!t]
  \includegraphics[width=\columnwidth]{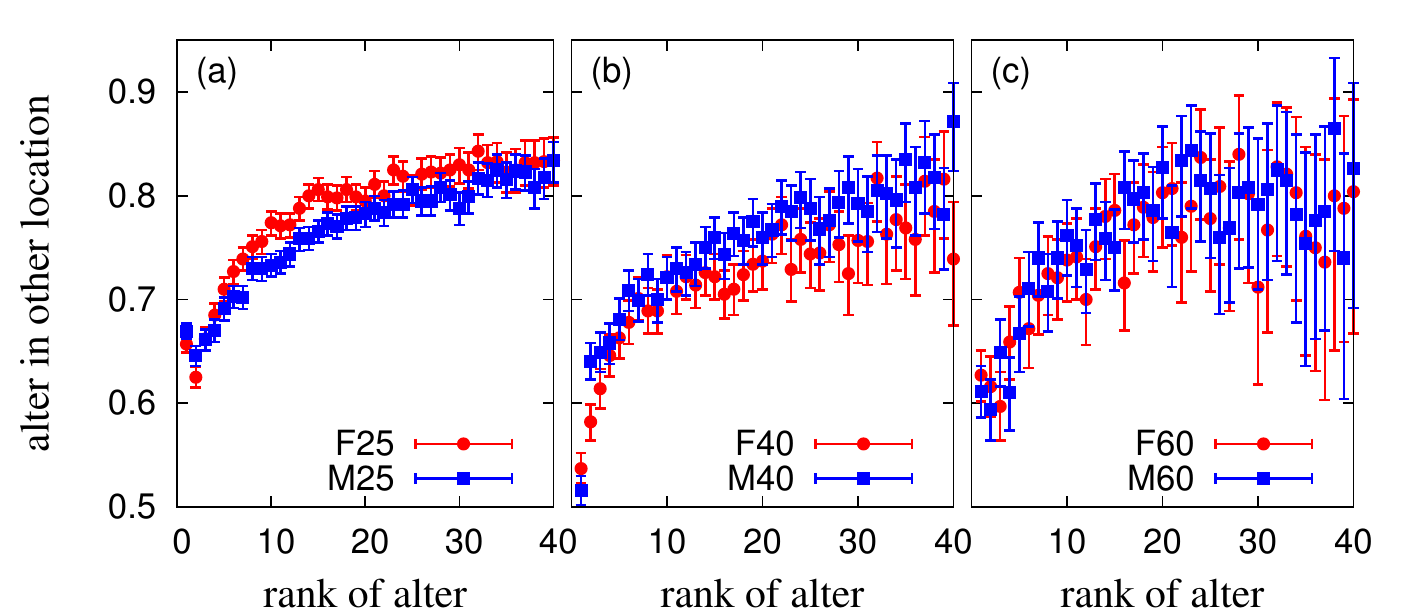}
  \caption{Geographic difference indices of alters who live in a different municipality to ego for several demographic groups of egos: F25 and M25 (a), F40 and M40 (b), and F60 and M60 (c). The rank of an alter is determined in terms of the number of calls to/from the ego. Error bars show the confidence interval with significance level $\alpha=0.05$.}
\label{fig:actLocMuni}
\end{figure}

We next extend our analysis from the highly-ranked alters to whole egocentric networks. Here the alters of an ego are ranked in a descending order according to the number of calls to/from the ego; this ranking typically reflects the emotional closeness to alters~\cite{Saramaki2014}. Figure~\ref{fig:egoNets} displays the geospatially embedded egocentric networks for four example egos; note the larger geographic spread for the younger egos. The geographic difference indices for groups of egos of specified age and sex and their ranked alters are shown in Fig.~\ref{fig:actLocMuni}. Overall, the geographic difference indices increase with increasing rank, implying that alters who are emotionally close also tend to be geographically close. This is in line with the results of Ref.~\cite{Lambiotte2008}, where it was shown that in a mobile communication network, on average, the probability that a tie belongs to a triangle decreases with the geographic distance of the two individuals. Since it is known that tie strength correlates positively with the density of triangles around the tie~\cite{Onnela2007}, this can be seen as indicative of strong ties being more frequently found between individuals who live close to one another. The results of Fig.~\ref{fig:actLocMuni} confirm this for egocentric networks.

Some sex differences can be observed in Fig.~\ref{fig:actLocMuni}. Among 25-year-olds, women are more likely to call alters who live in another municipality than are men (Fig.~\ref{fig:actLocMuni}(a)), but this pattern is reversed among 40-year-olds (Fig.~\ref{fig:actLocMuni}(b)). This distinction is lost among 60-year-olds (Fig.~\ref{fig:actLocMuni}(c)). In addition, it may be noted that 25-year-olds of both sexes are more likely to call and get called by alters who live further away than either 40-year-olds or 60-year-olds. This presumably reflects the fact that once married, egos are more likely to focus their attention on people who live closer, irrespective of whether these are other family members or friends.

\begin{figure*}[!h]
    \includegraphics[width=\columnwidth]{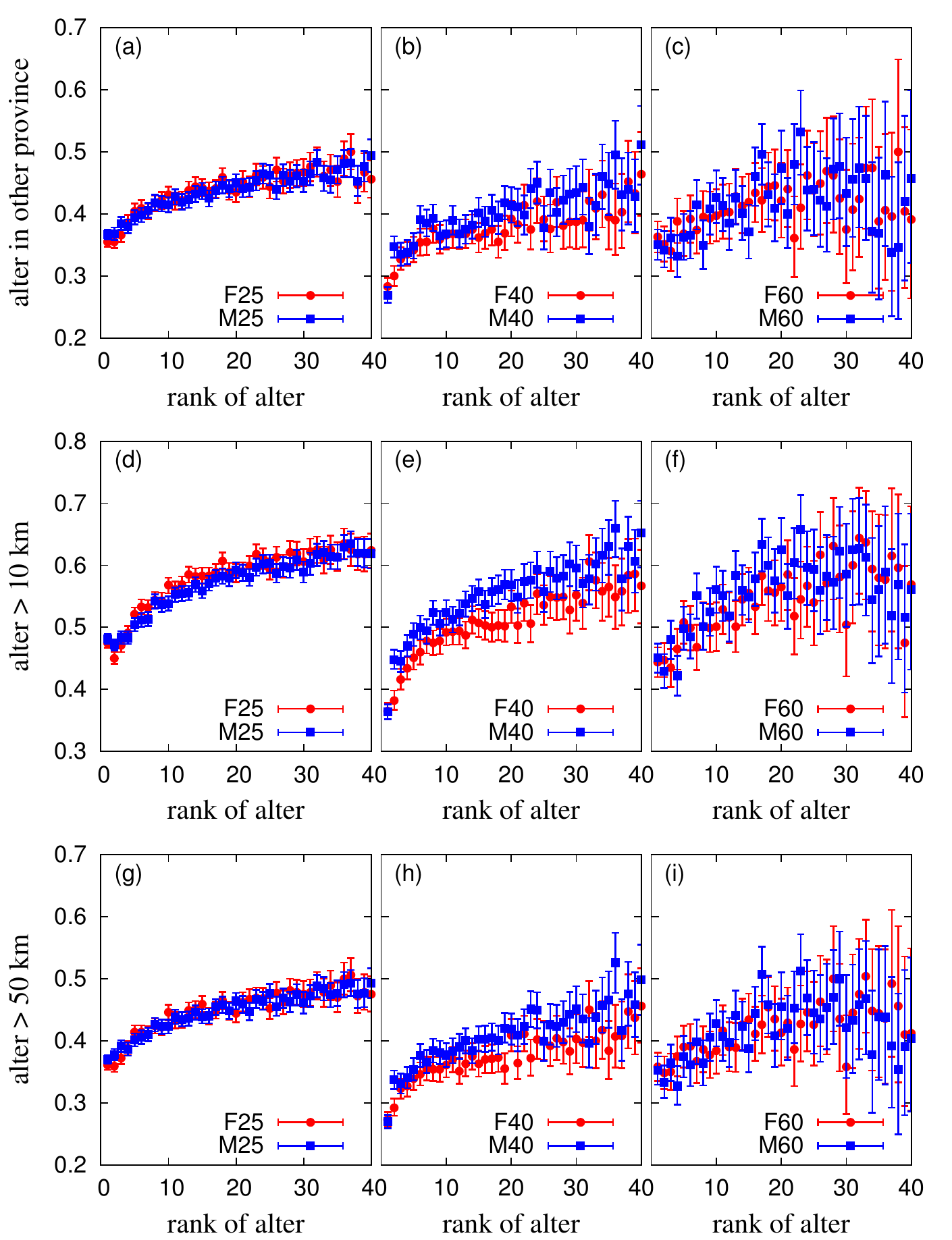}
  \caption{Geographic difference indices as functions of ranks of alters for several demographic groups of egos: F25 and M25 (left), F40 and M40 (center), and F60 and M60 (right). The rank of alter is determined in terms of the number of calls to/from the ego. Here the geographic difference indices are defined as the fraction of ego-alter pairs who live in different provinces (top), further than 10 km (middle), and further than 50 km (bottom). Error bars show the confidence interval with significance level $\alpha=0.05$.}
    \label{fig:actAll_error}
\end{figure*}

We have repeated all the above analyses by calculating the geographic difference indices based on provinces instead of municipalities, and with distance thresholds of 10 km and 50 km. The overall patterns are not affected by the choice of geographic difference index (see Supplementary Figs.~\ref{fig:allDiffLoc_error}, \ref{fig:groupAll_error}, and~\ref{fig:actAll_error}).

\section{Discussion}

We have analysed a large-scale mobile phone dataset to study the geographic correlations of emotionally close human relationships in the life-course framework. For this, we have made a number of assumptions: (i) Important relationships of mobile phone users are captured by call records, and the emotional closeness of a relationship is reflected in the strength of communication. (ii) The highly-ranked alters of the same age group as egos and of opposite sex than egos are likely to be egos' partners or spouses, while highly-ranked alters whose age is one generation apart from egos' age are likely to be egos' children or parents, irrespective of sex. (iii) The life-course migration patterns are reflected in the age- and sex-dependent geographic correlations of users. Based on these assumptions, we find that young couples tend to live further apart from each other than old couples. Female top-ranked alters of the same age and sex group as egos tend to live further apart than male top-ranked alters of the same age and sex group as egos. In all cases, the geographic correlation of close relationships with age difference larger than 10 years shows similar patterns independent of both sexes of egos and their highly-ranked alters. These findings are consistent with the finer scale analysis considering age correlations between highly-ranked alters and their geographic difference indices.

In addition, we find that emotionally closer friends tend to live closer to the ego. This is an important finding because it speaks against the common notion that people only take the trouble to call those who live further away from them (i.e., those individuals they cannot easily see face-to-face). In other words, the phone, particularly mobile phone, is not an alternative way of contacting distant alters, but rather a supplementary mechanism for contacting those alters that one also sees face-to-face.

\section{Methods}

\subsection{Data preparation and filtering} 

We use the mobile phone call dataset from a single mobile service provider in a European country. The dataset spans the first 7 months of 2007 and contains around 1.9 billion calls between 33 million individual users. Among them, around 5.1 million users subscribed to the operator, called company users. For company users, we have demographic information like sex and age, as well as geographic information like zip code and the most common location. The zip code is based on billing information, and the most common location is available in terms of latitude and longitude, i.e., the coordinate of GSM tower where the user spent the most time. The most common location is not necessarily the same as the location by zip code. Since the most common location is more informative than the zip code for studying behavioural patterns, we determine a zip code for the municipality nearest to the most common location of each user, which replaces the zip code based on the billing information whenever possible. By using the latitude and longitude of each municipality, the geographic distance between municipalities can be also calculated. Finally, the number of users whose information on sex, age, and zip code is available in the dataset is 3.2 million, i.e., 1.4 million females and 1.8 million males.

We find that for most egos, the age distribution of their alters shows a peculiarly sharp peak at the ego's own age. 
Such peaks are most likely anomalies due to multiple phones being registered to a single individual but used by his or her family members. In order to filter these anomalies, we assume that the number of alters of the same age as egos is equal to the bigger of numbers of alters who are one year younger or older than egos.

\subsection{Error estimation}

Finite sample sizes lead to errors in geographic difference indices. For each point of geographic difference index, the number of ego-alter pairs and the number of such pairs satisfying the condition are given as $n$ and $m$, respectively. Here the condition can be such that the alter lives in a different municipality to ego. Since the condition can be either satisfied or not, meaning a binary variable, we introduce a fraction $x$ of pairs satisfying the condition. The conditional probability of finding $m$ such pairs among $n$ pairs is $p(x|n,m)=x^m(1-x)^{n-m}/B_1(m+1,n-m+1)$, where $B_z(a,b)=\int_0^z x^{a-1}(1-x)^{b-1}dx$ denotes an incomplete beta function. For a given significance level $\alpha$, one may estimate the confidence interval $(x_{\rm min},x_{\rm max})$. Here the value of $x_{\rm min}$ is determined such that the cumulative probability for $x\in (0,x_{\rm min})$ is equal to $\alpha/2$, i.e.,
\begin{equation}
  \frac{B_{x_{\rm min}}(m+1,n-m+1)}{B_1(m+1,n-m+1)}=\frac{\alpha}{2}.
\end{equation}
From this equation, $x_{\rm min}$ is obtained as the inverse incomplete beta function. $x_{\rm max}$ is similarly obtained from
\begin{equation}
  \frac{B_{x_{\rm max}}(m+1,n-m+1)}{B_1(m+1,n-m+1)}=1-\frac{\alpha}{2}.
\end{equation}
Then, $(x_{\rm min},x_{\rm max})$ defines the confidence interval. These calculations are repeated for all points of geographic difference index.

\section*{Acknowledgements}

We thank A.-L.~Barab\'asi for the data used in this research. Financial supports by the Aalto University postdoctoral programme (H.-H.J.), by the Academy of Finland, project no. 260427 (J.S.), and by an ERC Advanced grant (R.I.M.D.) are gratefully acknowledged.





\bibliographystyle{apsrev}


\begin{figure*}[!h]
   \includegraphics[width=1.7\columnwidth]{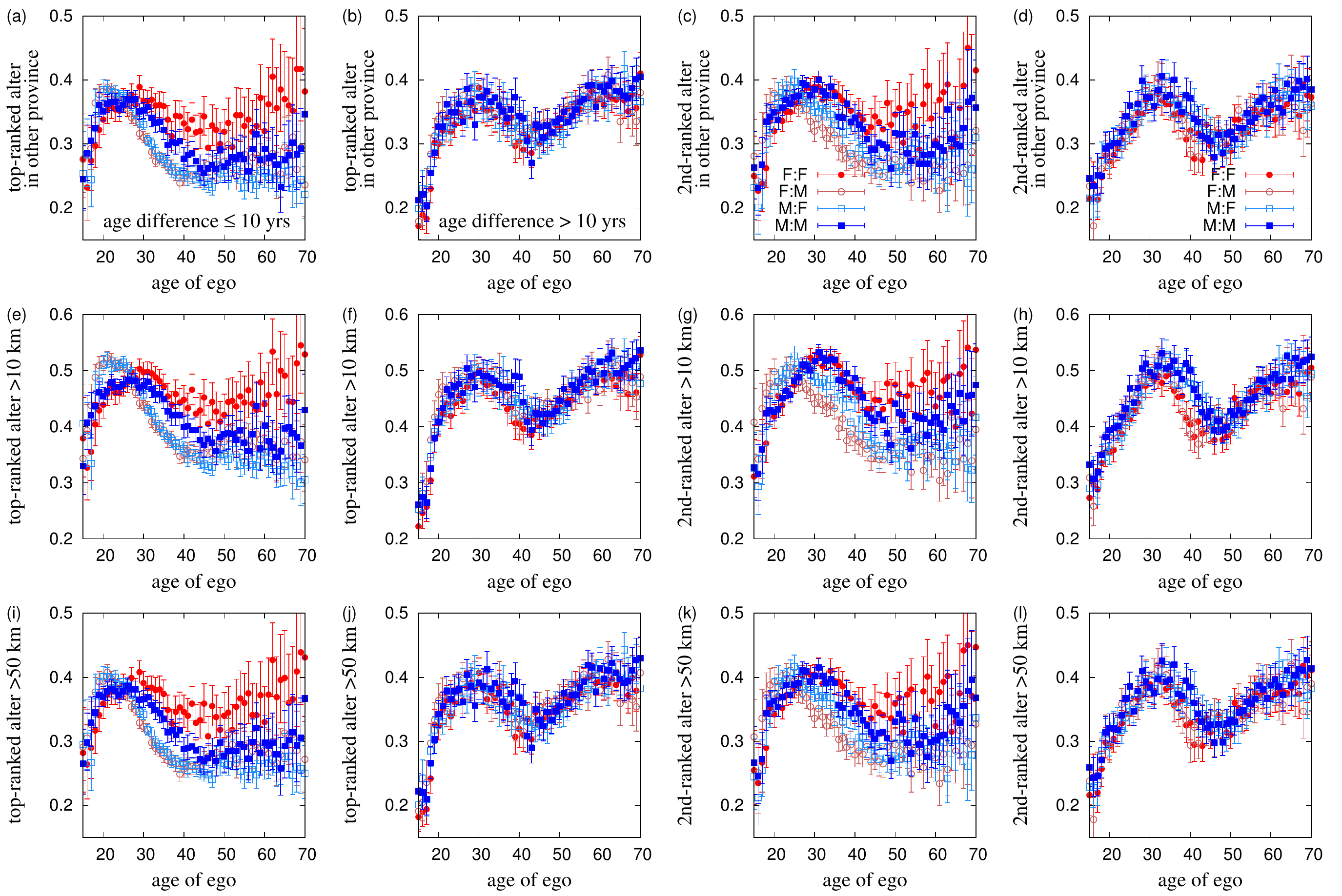}
   \caption{Geographic difference indices for top-ranked alters (left two columns) and 2nd-ranked alters (right two columns), where the geographic difference indices are defined as the fraction of ego-alter pairs who live in different provinces (top), further than 10 km (middle), and further than 50 km (bottom). Error bars show the confidence interval with significance level $\alpha=0.05$.}
   \label{fig:allDiffLoc_error}
\end{figure*}

\begin{figure*}[!h]
   \includegraphics[width=1.7\columnwidth]{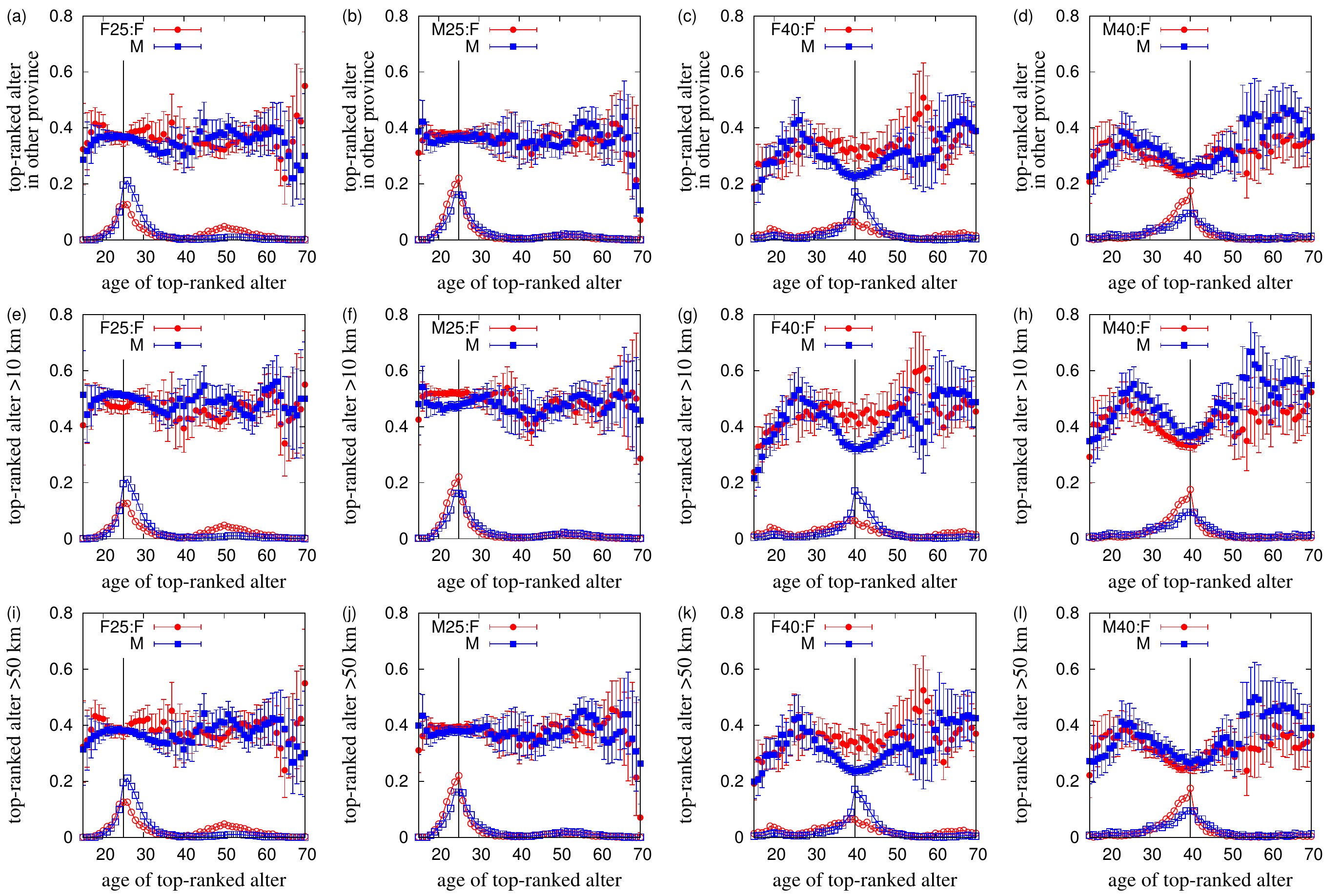}
    \caption{Geographic difference indices for top-ranked alters of ego groups F25, M25, F40, and M40 (from left to right), where the geographic difference indices are defined as the fraction of ego-alter pairs who live in different provinces (top), further than 10 km (middle), and further than 50 km (bottom). Error bars show the confidence interval with significance level $\alpha=0.05$. In each panel, lower curves with empty symbols represent age distributions of top-ranked alters, which are magnified $1.5$ times for clearer visualization. Vertical black lines at the ego age are added for guiding the eye.} 
    \label{fig:groupAll_error}
\end{figure*}

\end{document}